# Collagenase Nanocapsules: A Approach for Fibrosis Treatment.

*María Rocío Villegas[a,b], Alejandro Baeza [a,b]\*, Alicia Usategui [c], Pablo L Ortiz-Romero [d], José L. Pablos [c] and María Vallet-Regí [a,b]\**

a. Departamento de Química Inorgánica y Bioinorgánica, Facultad de Farmacia, Universidad Complutense de Madrid, 28040 Madrid, Spain.

b. Networking Research Center on Bioengineering, Biomaterials and Nanomedicine (CIBER-BBN), Spain.
E-mail: abaezaga@ucm.es ; vallet@ucm.es.

c. Servicio de Reumatología, Instituto de Investigación Hospital 12 de Octubre (I+12 Medical School), Universidad Complutense de Madrid, Spain.

d. Servicio de Dermatología, Instituto de Investigación Hospital 12 de Octubre (I+12 Medical School), Universidad Complutense de Madrid, Spain.

**Abstract**

Fibrosis is a common lesion in different pathologic diseases and is defined by the excessive accumulation of collagen. Different approaches have been used to treat different conditions characterized by fibrosis. FDA and EMA approved collagenase to treat palmar fibromatosis (Dupuyten's disease). EMA approved additionally its use in severe Peyronie's disease, but it has been used off label in other conditions.[1],[2] Approved treatment includes up to 3 (in palmar fibromatosis) or up to 8 (in penile fibromatosis) injections followed by finger extension or penile modelling procedures, typically causing severe pain. Frequently single injections are enough to treat palmar fibromatosis.[3] The need to inject repeatedly doses of this enzyme can be originated by by the labile nature of collagenase which exhibits a complete activity loss after short periods of time. Herein, a novel strategy to manage this enzyme based on the synthesis of polymeric nanocapsules which contains collagenase housed within their matrix is presented. These nanocapsules have been engineered for achieving a gradual release of the encapsulated enzyme for longer times which can be up to ten days. The efficacy of these nanocapsules have been tested in murine model of local dermal fibrosis yielding higher fibrosis reduction in comparison with the injection of free enzyme which represent a significant improvement over conventional therapy.

# 1. Introduction

Collagen is the most abundant protein in mammalian bodies (around of 30% of all proteins) and is a major structural component in the extracellular matrix (ECM).[4] There are different diseases which occur with an excess of collagen. The overproduction of collagen is caused by fibroblast hyperproliferation or by an unbalanced remodelation of ECM, among other reasons.[5] The accumulation of this higher content of collagen within a tissue can give rise to fibrosis, which compromises the tissue function causing physiological disorders and organ malfunction.[6],[7] Fibrosis is present in multiple conditions. Just to mention some of them, in superficial fibromatosis such as Dupuytren's disease, which consists in formation of a fibrotic cord in the hand, leading to hand dysfunction. In Peyronie's disease, penile involvement causes deviation making difficult or even impossible sexual relationships. Keloids or hyperplastic scars are consequences of abnormal scarring. Some autoimmune diseases as scleroderma characteristically present fibrotic/sclerous plaques on localized areas (morphea), generalized on the skin or even systemic (e.g. pulmonar involvement with lung fibrosis and high mortality). Fibromatosis have been traditionally treated with surgery in order to remove the collagen deposition.[6],[8] Unfortunately, surgical procedures often carry important complications producing serious adverse effects in many cases. In addition, the invasive nature of this type of interventions avoids the general application of surgical treatments, especially in fragile patients due to long post-operative recovery.[9] The high prevalence of associated morbidity in surgical treatments has led to the development of other type of treatments such as radiotherapy, oral administration of vitamin E (Tocopherol) or injection of corticosteroids, among others. However, these non-surgical treatments have not provided significant improvement in many cases.[4],[10],[11]

The administration of injections of Collagenase *Clostridium Histolyticum* (CCH) has emerged as one promising nonsurgical alternative for the treatment of fibrotic diseases because it is minimally invasive, cost-effective and presents lower complications and adverse side effects. [7],[9],[12] CCH is a matrix metalloproteinase mixture of two synergistic microbial collagenases, AUXI and AUX II, which act in combination and cleave the peptide bond of the collagen fibers between repeated sequences of Glycine-Proline-X (being X hydroxyproline or proline in the most effective cases).[4] Collagenase is able to lyse type 1 and type 3 collagen which are the which are the most abundant collagen types in fibrotic diseases. This enzyme shows high specificity for collagen fibrils whereas it does not produce any remarkable alteration in elastic fibers, vascular smooth muscle and axonal myelin sheaths.[10] Thus, this enzyme has been tested in the treatment of a widespread number of pathologies which course with an abnormal high regulation of formation and/or accumulation of collagen such as intervertebral disc herniaton,[13] vitrectomy,[14] burns,[15] wound healing process, keloid[16] and even in the treatment of solid tumors.[17] In this last application, the intratumoral injection of collagenase previous to the treatment with chemotherapeutic agents, both conventional drugs or nanoparticulated agents, allows a more homogeneous distribution of these therapeutic compounds along the tumoral tissue as a consequence of an improved diffusion of the molecules within a less dense extracellular matrix.[18]

Due to its probed effectiveness in some of these pathologies, CCH has been approved as enzymatic treatment by Food and Drug Administration in USA since 2 February 2010 and by European Agency for the Evaluation of Medical Products (EMEA) in January 2011, and it has been marketed under the name of Xiaflex (Endo Pharmaceuticals, Inc) for the treatment of Dupuytren, a palmar fibromatosis[4] and in December of 2013, for the treatment of Peyronie, a fibrosis that affect to tunica albuginea of the penis.[19] Label recommendations for treatment of Dupuytren's disease include up to three injections followed by finger extension 24-48h after injection, an usually painful procedure. However, majority of cases need only one injection on the fibrotic lesion.[3] Peyronie's disesase requires up to 8 collagenase injections followed by penile remodeling.[20] The need to apply repeated administration could be due to the low physico-chemical stability of collagenase which suffers rapid proteolysis when is exposed to the proteases present in tissues and also to relatively rapid denaturalization by oxidation or hydrolysis, among others. Thus, collagenase solved in physiological pH lost practically all its activity after 24 hours, which indicate the labile nature of this macromolecule. Due to its scarce life time, enzimatic activity is rapidly lost, and efficacy is limited. Repeated injections followed by finger extension or penile remodelation are time consuming, very painful for patients and increase the risk of adverse events. As example, it has been reported as relatively frequent the corporal ruptures of penis five days after the second injection of each cycle.[21]

In view of these data, is reasonable to think that injecting collagenase with long life time could increase efficacy, reduce the number of injections spare working time from health professionals and decrease the apparition of adverse events. Several alternatives have been evaluated to prolong the circulation times of these proteins, such as their transport immobilized on nanocarriers surfaces[22] or their encapsulation within polymeric shells.[17] However, these type of alternatives do not show a sustained release of the enzyme for long times. Herein, a novel polymeric nanocapsules which contains collagenases inside and capable to release the enzyme with a controlled kinetic achieving a prolonged and sustained effect over time is reported. Moreover, these nanocapsules protect the housed collagenase against external aggressions maintaining its catalytic activity of during long time. The efficacy of these capsules has been tested using a mouse model of localized dermal fibrosis (scleroderma) induced by repeated dermal injection of bleomycin yielding a significant reduction of the fibrotic lesion compared to the conventional administration of free collagenase. These results could pave the way for the clinical use of these nanocapsules to treat localized fibrotic diseases. As an additional advantage, the highly versatile nature of the presented methodology can be easily applied for the encapsulation of different proteins providing interesting alternatives of protein-based diseases.

## 2. Experimental Section
### 2.1. Materials

Collagenase Type I (Life Technologies); Acrylamide (Fluka); 2-Aminoethyl methacrylate hydrochloride (Sigma Aldrich); Ethylene glycol dimethacrylate (Sigma Aldrich); bysmethacrylamide (MBA) (Sigma Aldrich), Ammonium persulfate (Sigma

Aldrich); N, N,N',N'-Tetramethylethylenediamine (Sigma Aldrich); Amicon®Ultra-2mL Centrifugal Filters Ultracel®- 10K (Millipore); EnChek®Gelatinase/Collagenase Assay Kit (Life Technologies); 10X PBS Buffer pH=7.4 (Ambion)

## 2.2. Instrumental section

The hydrodynamic size of protein capsules was measured by means of a Zetasizer Nano ZS (Malvern Instruments) equipped with a 633 nm "red" laser. Transmission Electron Microscopy (TEM) was carried out with a JEOL TEM 3000 instruments operated at 300kV, equipped with a CCD camera. Sample preparation was performed by dispersing in distilled water and subsequent deposition onto carbon coated copper grids. A solution of 1% of phosphotungstic acid (PTA) pH 7.0 was employed as staining agent in order to visualize the protein capsules. Fluorescence was measured with Synergy 4, power supply for Biotek Laboratory Instrument 100-240VAC, 50/60Hz, 250W.

## 2.3. Synthesis of collagenase nanocapsules

Firstly, the reaction buffer $NaHCO_3$ (0.01 M, pH 8.5) was deoxygenated by freeze-vacuum-$N_2$ cycles. Then, Collagenase ($3.1 \times 10^{-5}$ mmol) was dissolved in 1ml of deoxygenated buffer. In an vial, acrylamide (AA), 2- aminoethyl metacrylate hydrochloride (Am), ethylene glycol dimetacrylate (EG) and bismethacrylamide (MBA) which proportions require in each case were dissolved in 1 mL of deoxygenated buffer and the monomers solution were added to the protein solution. This mixture was stirred at 300 rpm for 10 min under nitrogen atmosphere at room temperature. Then, 0.013 mmol of ammonium persulfate and 0.02 mmol of N,N,N′,N′-tetramethyl ethylenediamine (TMDA) dissolved in 1 mL of the deoxygenated buffer were added. The solution was stirred at 300 rpm for 90 min at room temperature under inert atmosphere. Next, the encapsulated enzyme was purified by centrifugal separation with 10 KDa cut-off filters (AMICON Ultra-2 mL 10 KDa) and washed three times with NaHCO3 buffer (0.01 M pH 8.5). The capsules of collagenase were preserved at 4 °C.

**2.4. Enzymatic activity assay**

The samples were incubated using the same collagenase concentration (measured by absorbance at 280nm) in a phosphate buffer with physiological pH=7.4 and at 37 ºC under stirring. The encapsulation process do not affect to absorbance since the monomers and polymer created do not have absorbance at this wavelength. At certain times, an aliquot of this solution was removed and its enzymatic activity was evaluated using the commercial kit (EnChekGelatinase/Collagenase Assay Kit). The kit provides a fluorimetric method and its protocol consists in mixing the collagenase sample with commercial buffer, which content calcium and fluorescent labeled gelatin. The fluorescence signal is quenched until gelatin digestion. Then, the enzymatic activity of the collagenase sample was determined by fluorescence release.

**2.5 Hydrolysis assay**

Collagenase nanocapsules were incubated in a phosphate buffer with physiological pH=7.4 and at 37 ºC under stirring. At certain times an aliquot was remove from the solution and staining with phosphotungstic acid 1% and observed by Transmission Electronic Microscopy (TEM).

**2.5. Mouse model of scleroderma**

Female 6-week-old C3H/HeNHse mice were purchased from Envigo (Valencia, Spain). Dermal fibrosis was induced by subcutaneous injections of 100 μg of bleomycin (1 mg·ml$^{-1}$, Mylan Pharmaceuticals, Barcelona, Spain) or 0.9% saline control into the shaved back skin every day for 4 weeks as previously described.[23] To analyze the effect of collagenase administration on this model after fibrosis induction, groups of 10 mice were given 300 micrograms of collagenase (Life Technologies), either in a single subcutaneous injection or encapsulated, in the bleomycin injected skin area and maintained for 10 days. In order to ensure that the amount of injected free and encapsulated collagenase is the same, the total protein content was determined by absorbance at 280nm. The study was approved by Animal Care and Use Committee of Hospital 12 de Octubre with protocol reference number PROEX 407/15 and carried out in accordance with the institutional guidelines.

Treated skin was harvested and paraffin embedded for histological evaluation of the collagen dermal area by using the Masson's trichrome staining kit (Sigma-Aldrich, St. Louis, USA). Masson's stained full thickness skin sections were photographed and digitalized using an AxioCam ERc 5S camera and ZEN lite 2012 software (Zeiss, Jena, Germany). Collagen blue-stained fractional area was quantified using ImageJ software (http://rsb.info.nih.gov/ij). The collagen area of each individual mouse was calculated as the mean area of several (3-5) sections of the central part of the excised skin piece.

## 2.6. Statistical analyses

Data were analyzed using Prism software (GraphPad Software, San Diego, CA, USA). Results are expressed as mean±SD and quantitative data were analyzed by Mann–Whitney U test. P values <0.05 were considered significant.

## 3. Results and Discussion

In this work, a protective polymer capsule is designed around the collagenase whose function is to release the house enzyme in controlled a sustained manner during large periods of time. In this way, the polymeric nanocapsule act as collagenase reservoir presenting a slow and tunable kinetic release capable to provide a prolonged effect, avoiding the need of important and repeat doses.

Protein nanocapsules have been widely studied in order to obtain a controllable delivery system which protect the cargo at the same time. In this sense, the development of protein carried based in its encapsulation within of polymeric mesh have received an increasing attention in the last few years.[24] These type of system has been demonstrated its capacity to trigger the protein release in response to different stimuli, as for example, enzymes present in the target site, according to local environment or specific cellular events.[25] In addition, these protein nanocapsules have been designed to achieve a sequential release of multiple protein, so that the release of the first protein triggers the release of the rest in a tandem sequence.[26] As added value, polymeric nanocapsules suppose a protective coating for the hosted enzyme against external insults that can compromise its structure and, therefore, its catalytic activity.[17]

Thus, with the objective to obtain a collagenase nanocapsules with a sustained release, a polymeric mesh was designed to coat the protein. Our nanocapsules have been made by free radical polymerization of two types of monomers, a neutral monomer, Acrylamide (AA), and 2-aminoethylmethacrylate hydrochloride (Am). AA is widely used as a structural monomer whereas Am present the capacity to adsorb around of negative proteins by electrostatic affinity allows due to its positive charge in aqueous phase. Thus, the enrichment of monomers around the proteins is favored by electrostatic bonding. Additionally, once the nanocapsule is formed, the presence of these amino groups on the surface avoids the capsule aggregation by repulsion charge enhancing the colloidal stability of the system. Finally, the ethyleneglycol dimethacrylate (EG) was chosen as a degradable crosslinker to form the polymeric shell.[27]

First, the monomers and the protein are put in contact in an aqueous solution free of oxygen. The protein and monomers mixture was keep during 10 min under stirring in order to allow that the monomers were adsorbed on the negative surface protein via intermolecular interactions. Then the protein surface could be enriched in monomers in such a way the polymerization happens around the protein. Oxygen, which can cease the radical polymerization, is removed from the buffer reaction previous to the addition of radical initiators. This first stage is followed by a polymerization step. The polymerization initiates by the addition of ammonium persulfate (AP) in the presence of *N,N,N',N'*-tetramethylethylenediamine (TMDA) which forms free radicals of oxygen in aqueous solution at room temperature by a basic catalysis mechanism. AP as the initiator of radical creation and are catalyzed by TMDA which promote the decomposition of AP into free radical decreasing the activation energy of polymerization.[28] In this way, the addition of TMDA allow the free radical polymerization process happens a room temperature. The free radical process leads to polymerization of acroyl groups of monomers and crosslinker yielding polymeric mesh that coat the protein (**Scheme 1**). The mixture was kept at room temperature during 90 min and collagenase nanocapsules were isolated by centrifugal separation with 10KDa cut-off filters in order to remove the monomer excess. Collagenase acts on large substrates (collagen fibers) and therefore it is necessary its release in order to facilitate its catalytic activity. In order to provide a controlled release mechanism, EG was chosen as pH responsive crosslinker. Thus, the obtained nanocapsules would be degradable by hydrolysis. The collagenase release mechanism is showed in Scheme 2. In aqueous medium, water molecules act as nucleophiles attacking the carbonyl group of the crosslinker which results in the hydrolysis of the ester bonds. This crosslinker degradation is catalyzed in acidic environments, due to an interaction of protons with the oxygen atom of the carbonyl group, increasing the polarity of the covalent bond and favoring the nucleophilic attack by water.

**3.1. Optimization of Collagenase Synthesis to Sustained Release**
**3.1.1. Previous work.**

Our research group recently reported the encapsulation of collagenase inside degradable polymeric nanocapsules using a ratio protein: monomers 1:2025 and AA/Am/EG monomer ratio of 7/6/2.[17] These conditions provided nanocapsules which release the hosted enzyme in less than 12h at physiological conditions. This supposes a high improvement of the stability of the enzyme and this relatively rapid release is suitable for certain clinical applications, in the case of fibrotic disease, longer release times is required as have been mentioned.

### 3.1.2. Strategies to achieve a sustained and prolonged release.

Based on the previous work, different strategies have been performed in order to achieve a prolonged and sustained release. In this way, is hoped the collagenase nanocapsules can act a depot of the proteolytic enzyme and improve the temporal window in this act enhanced its therapeutic effect. Two strategies have been developed in order to achieve this goal, first the degradable crosslinker proportion was reduced to try reduce the hydrolysis attack point. On the other hand, the obtaining of collagenase nanocapsules with sustained release was addressed from the other point of view, this is the addition of an additional crosslinker, non-degradable, in this way the polymeric mesh would turn more robust and the collagenase release rate would be decreasing.

### 3.1.2.1. Decrease in the proportion of degradable crosslinker

In order to decrease the hydrolysis rate of the collagenase nanocapsules the amount of degradable crosslinker was reduced to the half respect of previous work.[17] Protein to monomer ratio was maintained constant because a significant alteration in this ratio can lead to poor protein encapsulation. The capacity to provide a sustained enzymatic activity of these nanocapsules during time was evaluated incubating the nanocapsules in an aqueous solution at physiological pH and temperature under orbitalic soft stirring. Then, at certain times, an aliquot of this solution was removed and used to measure the enzymatic activity using commercial kit protocol. The nanocapsules retain more than 50% of the catalytic capacity initial during 2 days, which is a significant improvement in comparison with the free enzyme, which lost more than 80% of their capacity after 1 day. The encapsulation efficacy is also dependent of the facility of the monomers to be adsorbed on the protein surface and this capacity is related with the intermolecular interactions between both systems. In order to evaluate if the AA/Am monomer ratio could play a role in this process, two different monomer relations were also evaluated, AA/Am/EG 7/7/1 and 8/6/1, respectively. Neither of these cases provided a significant improvement in the catalytic activity during time, losing more than 50% of their capacity after 2 days, similar to the previous case (**Figure 1**).

The presence of crosslinker is strictly needed to obtain a dense and protective mesh around the protein. Then, although the higher decrease in the crosslinker ratio could be an option in order to extent the time release, it could result in an inappropriate capsule formation. For this reason, the possibility to reduce even more its amount was discarded and other strategy was addressed.

### 3.1.2.2. Introduction of non-degradable crosslinker

Instead of reducing the crosslinker, a novel strategy which consists in the introduction of an additional non-degradable crosslinker, bismethacrylamide (MBA) was employed. The introduction of this crosslinker should not alter the nanocapsule formation since the ratio crosslinker: monomer was maintained and could increase the degradation time of the nanocapsule allowing a sustained release at larger times. Although, the addition of non-degradable crosslinker can result in incomplete collagenase release, this strategy decreases the hydrolysable points of polymeric mesh without altering the crosslinker ratio. Thus, fixing the ratio protein:monomer=1:2025 and AA/Am/crosslinker=7/6/2 which had provided good encapsulation performance, different ratios of each type of crosslinker, degradable or not degradable, was studied. The addition of the new crosslinker did not altered the size and morphology of the nanocapsules in all cases. The results indicate that employing a 1:1 ratio between degradable and not degradable crosslinker, the system is able to maintain the enzymatic activity for really long times. This sample shows a sustained release around 50% of its initial activity lasting 10 days and maintaining the 15% of its activity at 12 days (**Figure 2**). Surprisingly, by unknown reasons, employing different proportions between the crosslinkers, the capacity to maintain the catalytic activity drops to values even lower than the previous systems which carry only the degradable crosslinker, both if the non-degradable linker is employed in higher amount and when is added in the lower ratio. In any case, the system which employ 1.1 ratio exhibit excellent properties. The hydrolysis of this sample was observed by TEM, in the micrographs obtained (**Figure 3**) can be observed as the collagenase nanocapsules progressively loss its integrity as is expected due to are degradable nanocapsules. Due to its sustained release sample with ratio EG/MBA=1/1 was chosen for in vivo evaluation in comparison with the current treatment.

Therefore, two strategies were evaluated in order to achieve a sustained and prolonged release of collagenase. Both approaches resulted in an increase of the collagenase release time respect of our previous work. In the first study, a decrease of degradable crosslinker in the collagenase nanocapsules synthesis was evaluated. This strategy was based on the idea that decreasing the degradable crosslinker ratio led to a reduction in the attack points by hydrolysis. Although the stability of the collagenase nanocapsules were higher than free enzyme, the results were not satisfactory because the collagenase release was not sustained losing almost completely the enzymatic activity after three days. The idea to keep reducing the crosslinker ratio was discarded since the presence of crosslinker is necessary in order to obtain a robust polymeric coating. Thus, the problem was addressed from other point of view; by the introduction of an additional non-degradable crosslinker. Despite the fact that the addition of non-degradable crosslinker presented a potential liability regarding with the possibility to led to an incomplete collagenase release, it allowed to decrease the hydrolysis-sensitive points of the nanocapsules without altering the crosslinker ratio. This approach yielded nanocapsules with excellent properties for the treatment fibrotic lessions, because they exhibited a sustained a prolonged collagenase release during up to 10 days. Once optimized the synthetic conditions for the formation of collagenase nanocapsules with desired properties, the suitability to treat fibrotic lesions with this nanodevice was evaluated in a mice model as proof of concept in comparison with the conventional free enzyme administration.

### 3.2. *In Vivo* Cytotoxicity Test of Collagenase Nanocapsules

To check the biocompatibility of the collagenases nanocapsules in vivo, a fixed amount of nanocapsules or free collagenase was injected subcutaneously into the shaved back skin of healthy mice. After 10 days, the animals were sacrificed and histological analysis of collagenase injected skin showed a normal structure of all skin layers and absence of inflammatory cell infiltration (**Figure 4**). Dermal collagen area and structure did not show changes compared to normal (uninjected) skin. (Normal skin *vs* free collagenase p=0.7, normal skin *vs* nanocapsules p=0.99, free collagenase *vs* nanocapsules p=0.628).

### 3.3. *In Vivo* Evaluation of the Efficacy of Collagenase Nanocapsules

In order to study the efficacy of collagenase nanocapsules on bleomycin-induced skin fibrosis, we evaluated changes in the fibrotic area induced by bleomycin for 4 weeks, after injection of either free collagenase or collagenase nanocapsules. We found a marked increase of the dermal collagen area of the dermis after bleomycin injection compared to control (saline-injected) group (**Figure 5**).

The fibrotic area, evaluated as the fractional collagen stained area (blue-stained area) of the dermis, was significantly reduced in mice injected with bleomycin and collagenase nanocapsules-treated compared to bleomycin mice (p<0.0001) and bleomycin/free collagenase-treated group (p=0.026). There were no significant differences between bleomycin/free collagenase and bleomycin groups (p=0.457; **Figure 5**). Therefore, administration of a single dose of encapsulated collagenase significant decreased the collagen area, whereas the same dose of collagenase without encapsulation (free collagenase) was insufficient to reduce the collagen area.

The fact that a single injection of collagenase nanocapsules results in significant collagen area decrease respect to free enzyme could suppose an important advance in the fibrotic lesions treatment. In future Works, preclinical studies should be organized in order to prepare clinical trials in humans

4. **Conclusion**

Herein, a novel collagenase nanocapsules capable of a sustained release and delivery of proteolytic enzyme has been described. These collagenase nanocapsules were designed to protect the enzymatic activity of the collagenase and release with a controllable kinetic during large and modulable periods of time. These polymeric nanocapsules suppose a delivery system of collagenase able to release the enzyme at physiological conditions during 10 days.

This type of release is highly beneficial in the enzymatic treatments, since a prolonged effect in the time of the required enzyme is obtained allowing the use of lower doses and/or reducing the number of injections necessary to achieve acceptable results. The high stability conferred by this coating composition coupled with the sustained release profile results in a promising advance in the stabilization of the enzyme and makes it possible to improve the current clinical treatment. It has been observed in fibrotic models, where the collagenase encapsulated showed a highly efficacy of degradation in comparation to free enzyme, which is the actual treatment. The application of this strategy would pave the way for the delivery of proteins able to achieve more sustained release, which would result an important advance to their clinical applications.

5. **Acknowledgements**

This work has been done thanks to the financial support provided by European Research Council (Advanced Grant VERDI; ERC-2015-AdG Proposal No. 694160) and the project MAT2015-64831-R. In addition, this work was supported by grants from the Instituto de Salud Carlos III (Ministerio de Economía y Competitividad, Spain) PI 12/439 and RIER (Red de Investigación en Inflamación y Enfermedades Reumáticas) to AU and JLP, and co-financed by FEDER (European Union).

# Figure of new document

**Scheme 1:** Synthesis of collagenase nanocapsule by free radical polymerization. AA, Am, EG, PA/TMDA are referred to acrylamide, 2-Aminoethylmethacrylate hydrochloride, ethylene glycol dimethacrylate, ammonium persulfate/ N, N, N´, N´, tetramethylenediamine respectively.

**Scheme 2:** Degradation mechanism of the collagense nanocapsules.

**Figure 1:** a) Study of Dynamic Light Scaterring (DLS) of collagenase nanocapsules with different AA/Am ratios. Study of stability of collagenase nanocapusules with different ratios AA/Am (Purple). Micrograph by transmission Electron Microscopy b) AA/Am/EG=7.5/6.5/1, c) AA/Am/EG=7/7/1 d) AA/Am/EG=8/6/1 (white scale bar correspond to 100nm). Study of free collagenase was represented in blue.

**Figure 2:** a) Study of Dynamic Light Scaterring (DLS) of collagenase nanocapsules with different EG/MBA ratios. Study of stability of collagenase nanocpasules with different ratios EG/MBA (Purple). Micrograph by transmission Electron Microscopy b) AA/Am/EG/MBA=7/6/1/1, c) AA/Am/EG/MBA=7/6/1.5/0.5 d) AA/Am/EG/MBA=7/6/0.5/1.5 (white scale bars correspond to 100nm). Study of free collagenase was represented in blue.

**Figure 3:** TEM micrographs of collagenase nanocapsules incubated at physiological conditions at different times.

**Figure 4:** Study of biocompatibility of nanocapsules. Mice received a single subcutaneous injection with a fixed amount of nanocapsules or free collagenase. Skin was stained by Masson´s trichrome (collagen fibers are stained in blue). Data are representative of one experiment with three-four mice per group. Bar 50µm.

**Figure 5:** Effect of collagenase nanocapsules in bleomycin induced skin fibrosis. C3H mice received daily subcutaneous injections of bleomycin and were treated with free collagenase or nanocapsules. Control mice were daily injected with saline. Fibrosis was quantified as the collagen area (blue-stained area as stained by Masson trichrome), expressed as a fraction of the total skin area. Data are representative of two independent experiments with 10 mice per group. Bar 50µm.

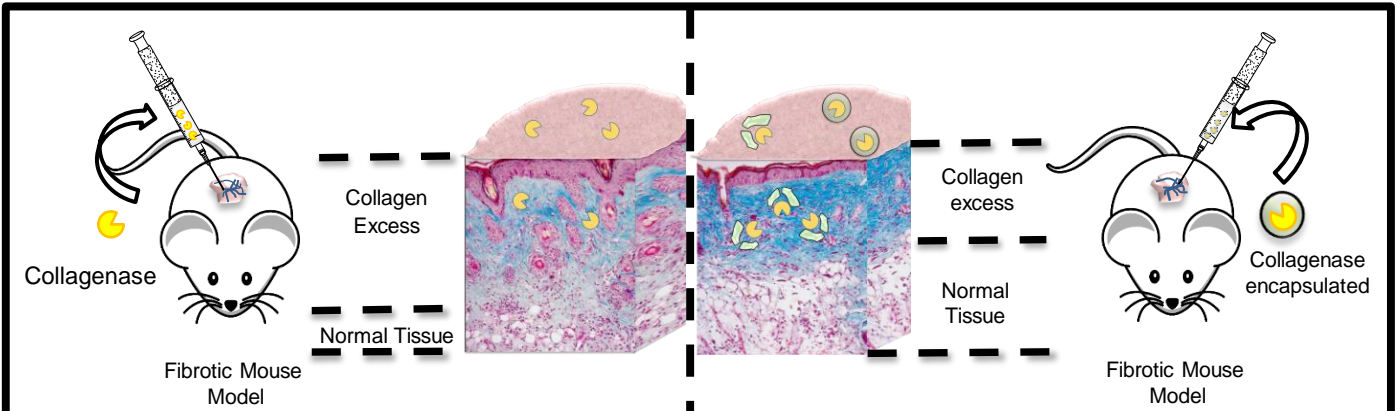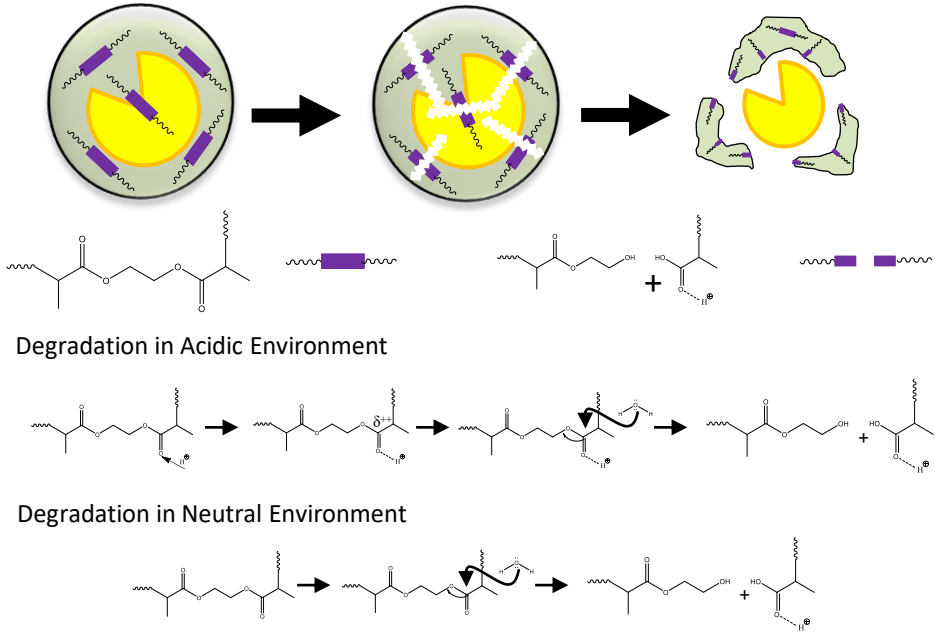

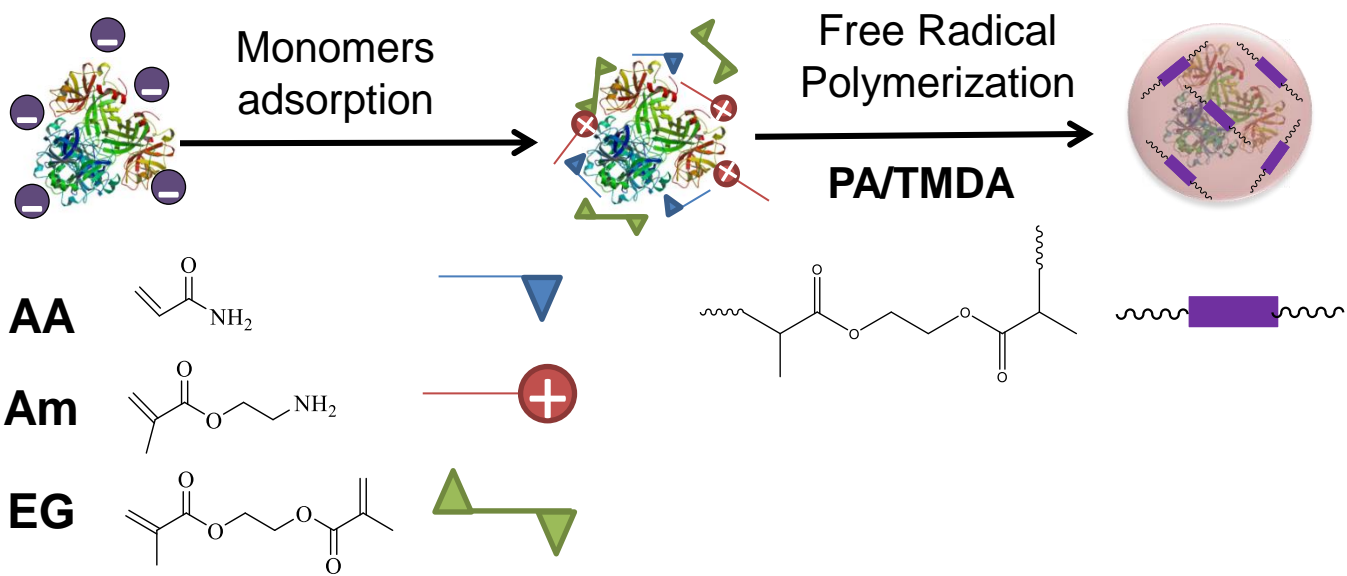

### Free radical initiation

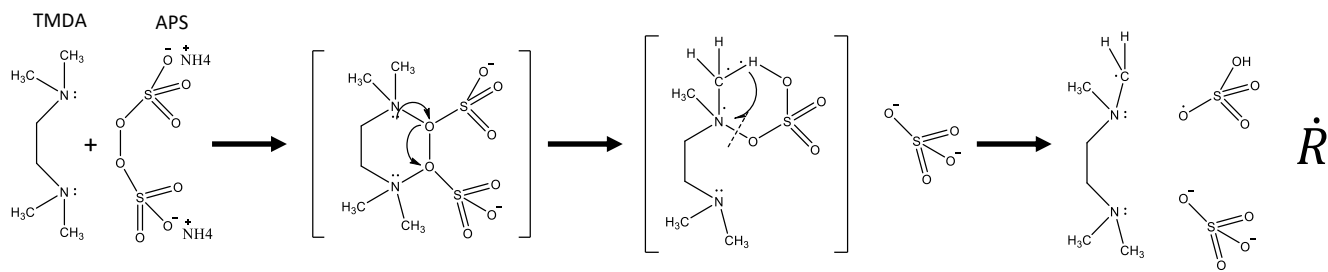

### Free radical polymerization

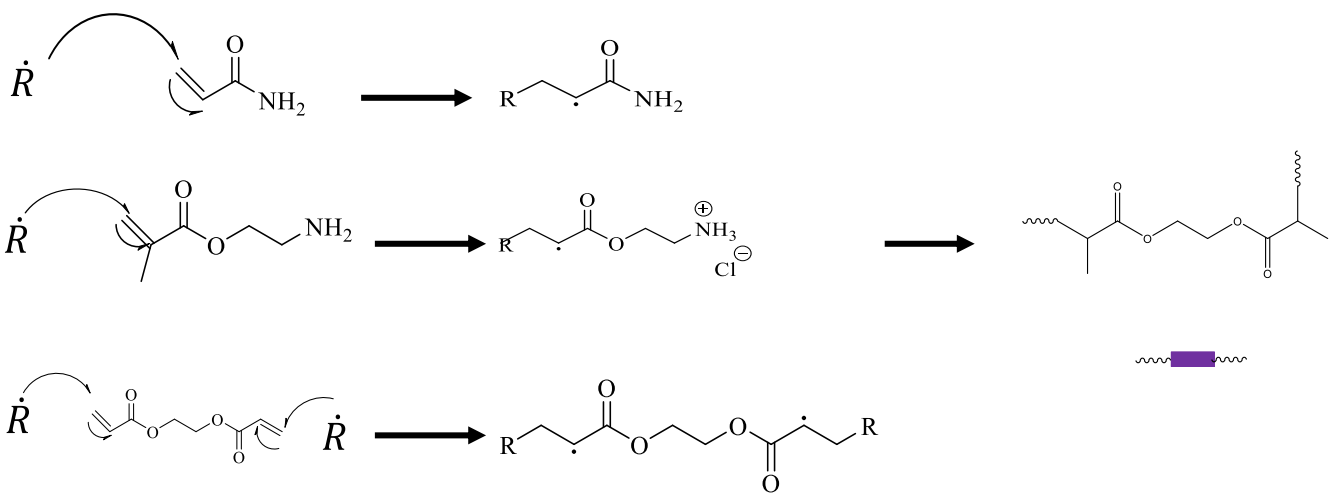

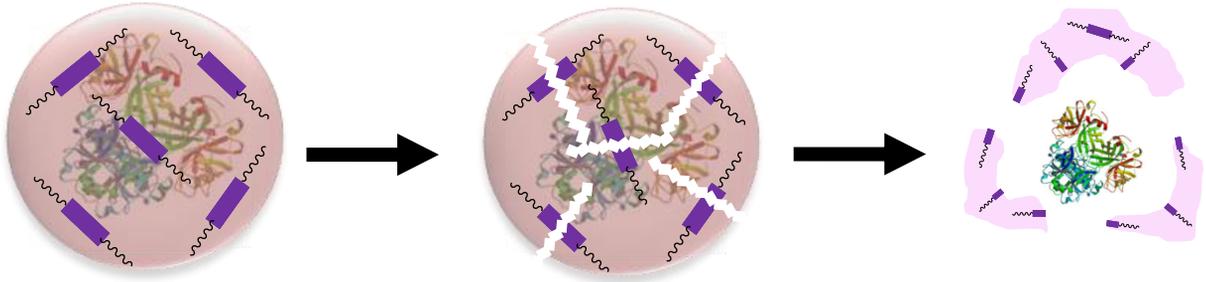

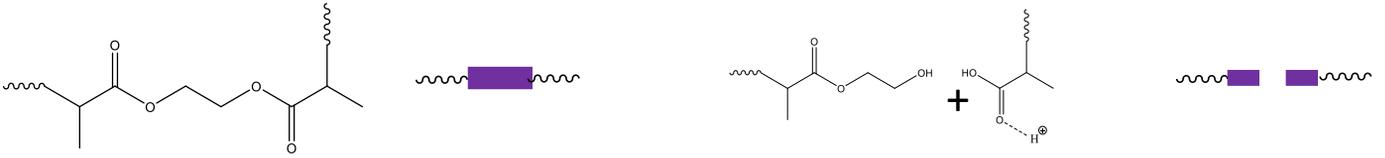

## Degradation in Acidic Environment

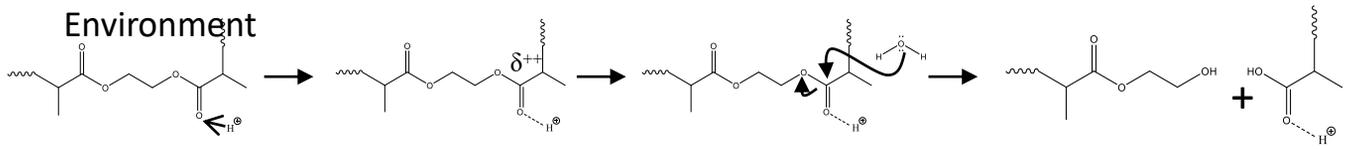

## Degradation in Neutral Environment

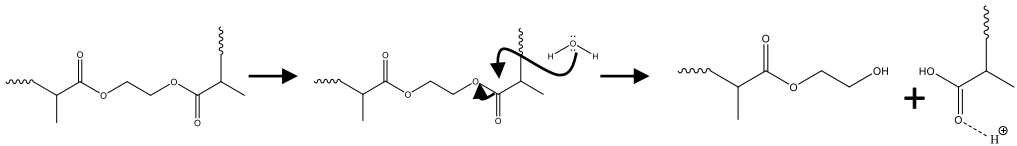

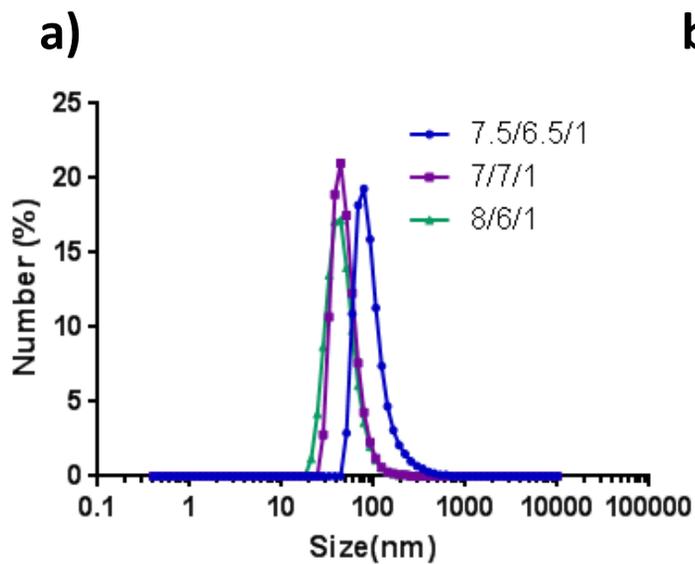
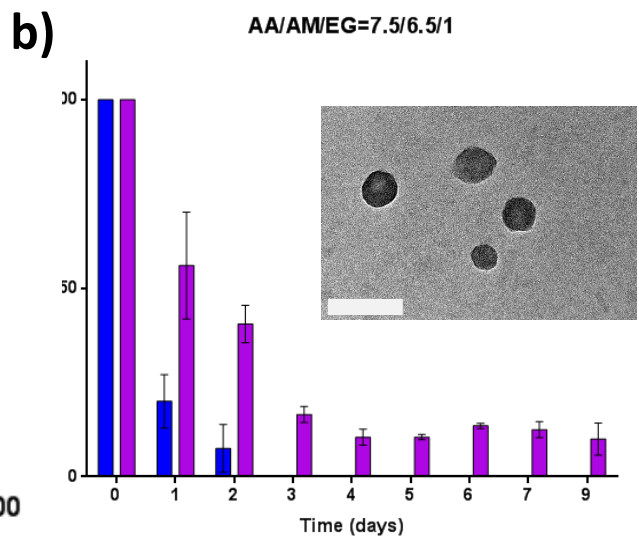
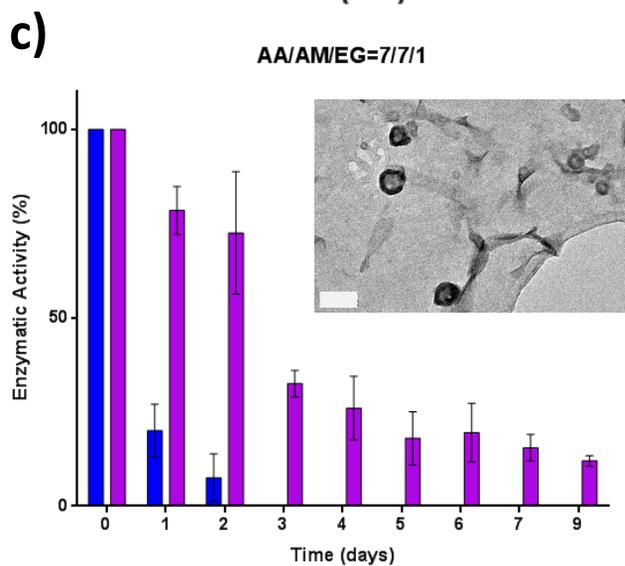
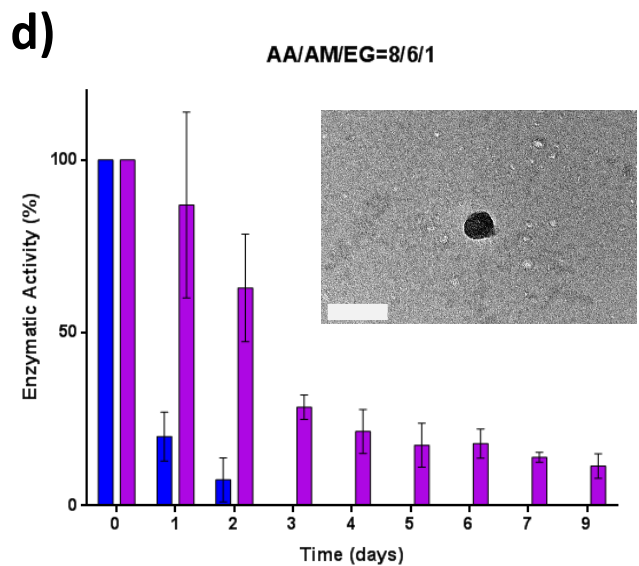

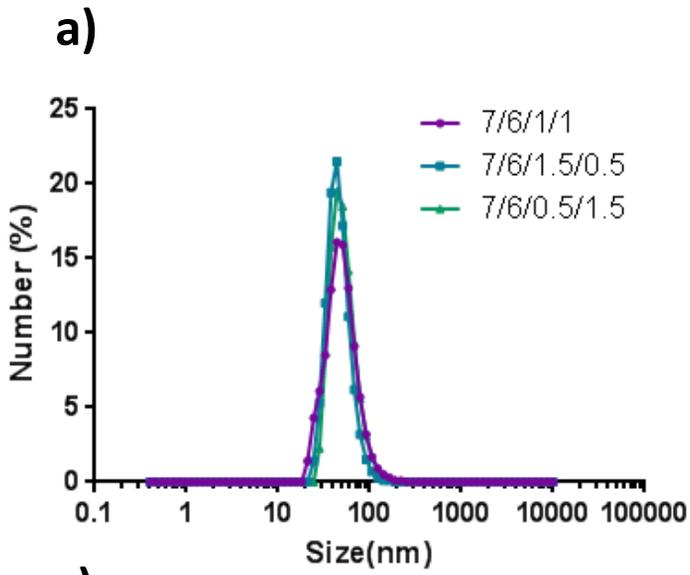
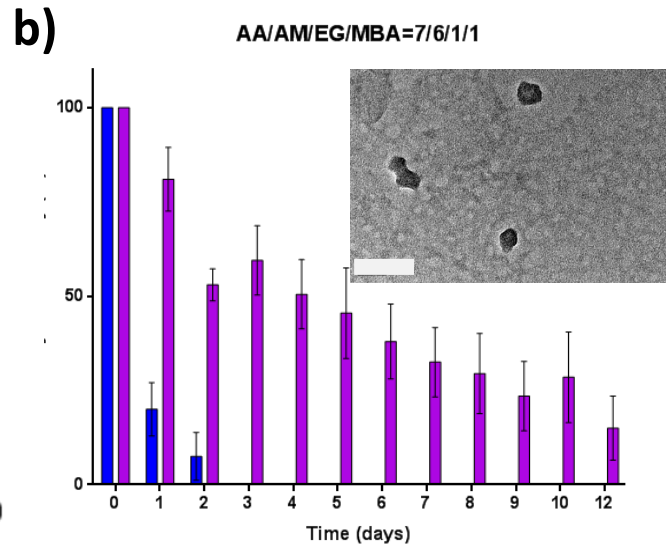
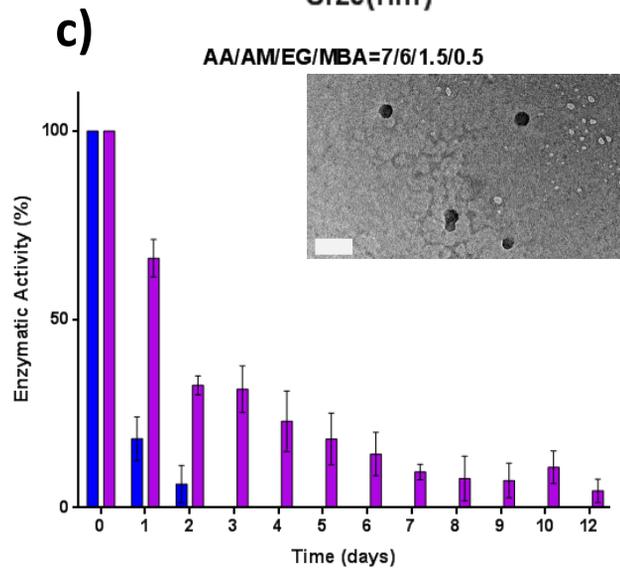
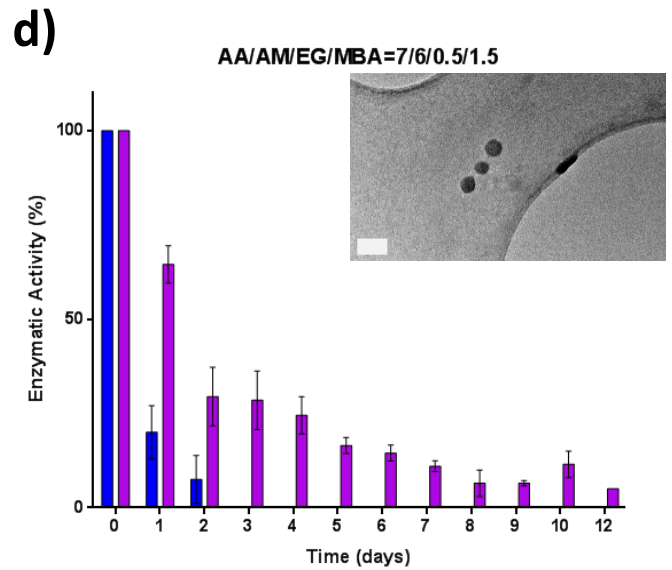

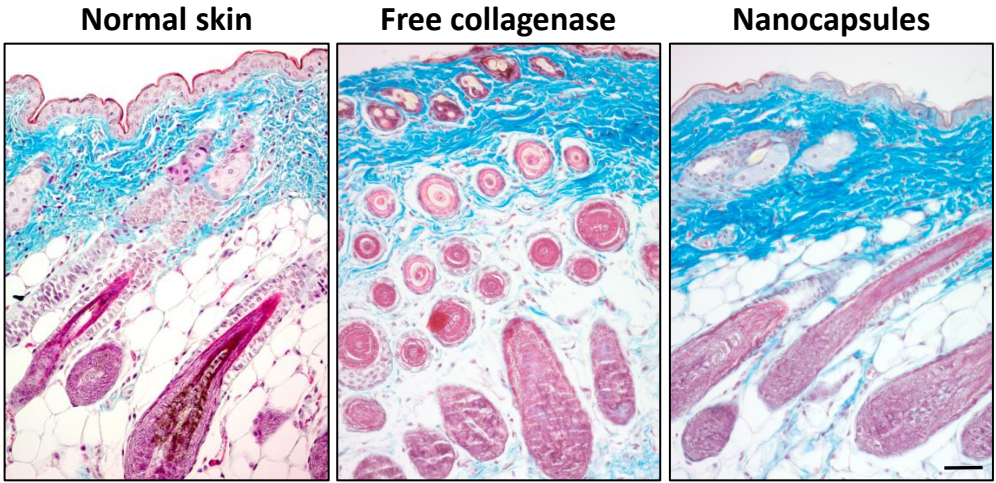

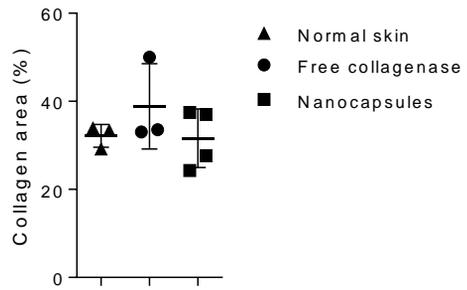

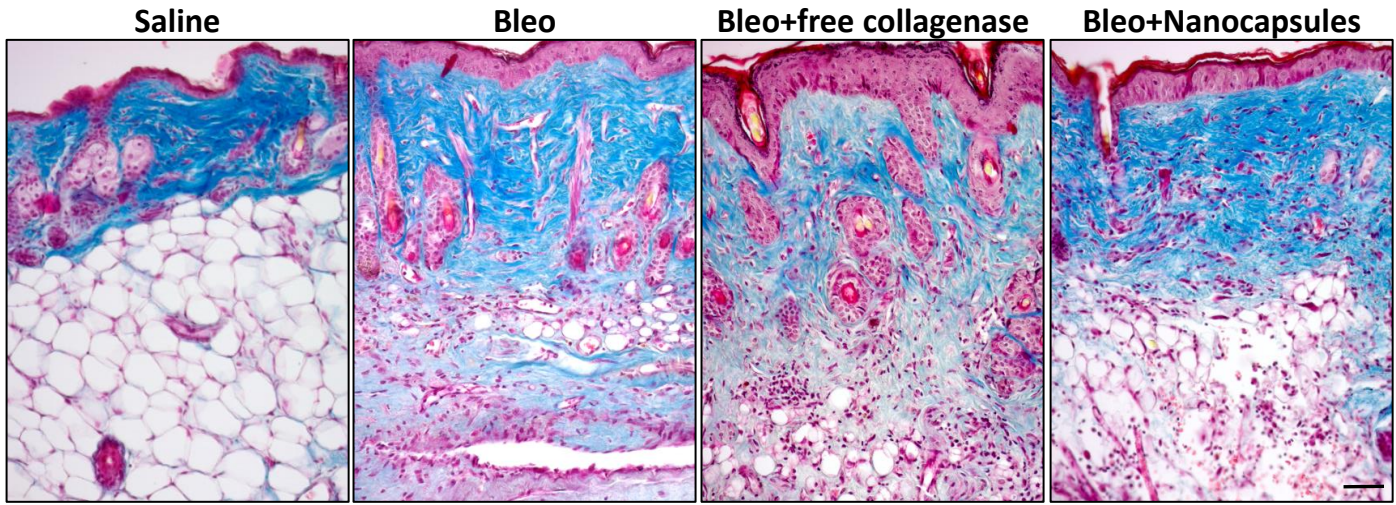

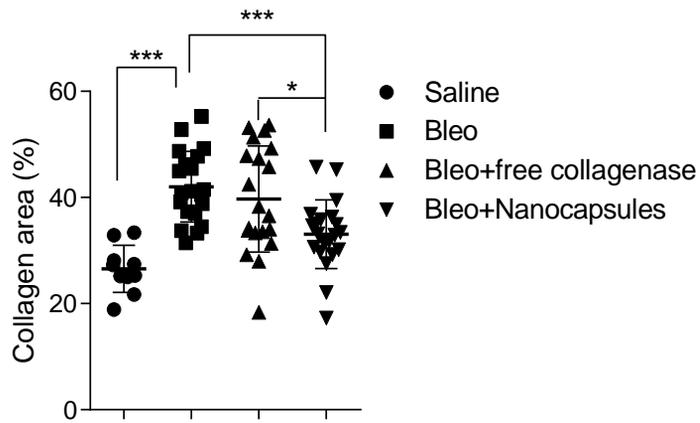

**Scheme 1:** Synthesis of collagenase nanocapsule by free radical polymerization.

**Scheme 2:** Degradation mechanism of the collagense nanocapsules..

**Figure 1:** a) Study of Dynamic Light Scaterring (DLS) of collagenase nanocapsules with different AA/Am ratios. Study of stability of collagenase nanocpasules with differents ratios AA/Am. Micrograph by transmission Electron Microscopy b) AA/Am/EG=7.5/6.5/1, c) AA/Am/EG=7/7/1 d) AA/Am/EG=8/6/1 (white scale bar correspond to 100nm). Study of free collagenase was represented in blue.

**Figure 2:** a) Study of Dynamic Light Scaterring (DLS) of collagenase nanocapsules with different EG/MBA ratios. Study of stability of collagenase nanocpasules with differents ratios EG/MBA. Micrograph by transmission Electron Microscopy b) AA/Am/EG/MBA=7/6/1/1, c) AA/Am/EG/MBA=7/6/1.5/0.5 d) AA/Am/EG/MBA=7/6/0.5/1.5 (white scale bars correspond to 100nm). Study of free collagenase was represented in blue.

**Figure 3:** Study of biocompatibility of nanocapsules. Mice received a single subcutaneous injection with a fixed amount of nanocapsules or free collagenase. Skin was stained by Masson´s trichrome (collagen fibers are stained in blue). Data are representative of one experiment with 5 mice per group. Bar 50μm.

**Figure 4:** Effect of collagenase nanocapsules in bleomycin induced skin fibrosis. C3H mice received daily subcutaneous injections of bleomycin and were treated with free collagenase or nanocapsules. Control mice were daily injected with saline. Fibrosis was quantified as the collagen area (blue-stained area as stained by Masson trichrome), expressed as a fraction of the total skin area. Data are representative of one of two independent experiments with 10 mice per group. Bar 50μm.